\documentclass[a4paper]{article}

\usepackage{INTERSPEECH2019}

\usepackage{url}
\usepackage{float}

\title{The Airbus Air Traffic Control speech recognition 2018 challenge: towards ATC automatic transcription and call sign detection}
\name{Thomas Pellegrini$^1$, J\'er\^ome Farinas$^1$, Estelle Delpech$^2$, Fran\c cois Lancelot$^2$}

\address{$^1$IRIT, Universit\'e Paul Sabatier, CNRS, Toulouse, France\\
$^2$Airbus, Toulouse, France}

\email{\{thomas.pellegrini,jerome.farinas\}@irit.fr\\\{estelle.e.delpech,francois.lancelot\}@airbus.com}

\begin{document}
%
\maketitle
\begin{abstract}
In this paper, we describe the outcomes of the challenge organized and run by Airbus and partners in 2018 on Air Traffic Control (ATC) speech recognition. The challenge consisted of two tasks applied to English ATC speech: 1) automatic speech-to-text transcription, 2) call sign detection (CSD). The registered participants were provided with 40 hours of speech along with manual transcriptions. Twenty-two teams submitted predictions on a five hour evaluation set. ATC speech processing is challenging for several reasons: high speech rate, foreign-accented speech with a great diversity of accents, noisy communication channels. The best ranked team achieved a 7.62\% Word Error Rate and a 82.41\% CSD F1-score. Transcribing pilots' speech was found to be twice as harder as controllers' speech. Remaining issues towards solving ATC ASR are also discussed in the paper.
\end{abstract}


\section{Introduction}
\label{sec:intro}

The recent advances in Automatic Speech Recognition (ASR) and Natural Language Understanding (NLU) technologies have opened the way to potential applications in the field of Air Traffic Control (ATC).

On the controllers' side, it is expected that these technologies will provide an alternative modality for controllers. As a matter of fact, controllers have to keep track of all the clearances they emit, this is nowadays made either by mouse input or by hand -- which generates a high workload for controllers. The ongoing research project MALORCA\footnote{\url{http://www.malorca-project.de/}}, for instance, aims at improving ASR models for providing assistance at different controller working positions.

On the pilots' side, ASR of ATC messages could also help decreasing pilots' cognitive workload. Indeed, pilots have to perform several cognitive tasks to handle spoken communications with the air traffic controllers:
\begin{itemize}
    \item constantly listening to the VHF (Very High Frequency) radio in case their \emph{call sign} (i.e. their aircraft's identifier) is called;
    \item understanding the controller message, even if pronounced with non-native accent and/or in noisy conditions;
    \item remembering complex and lengthy messages.
\end{itemize}

\begin{table*}
\centering
\caption{CTS speech (SWITCHBOARD) vs. ATC speech.} 
\label{tab:atc_vs_swb}
\vspace{2mm}
 \begin{tabular}{l | l | l}
   ~ & \bfseries{SWITCHBOARD speech} &  \bfseries{ATC speech} \\ \hline 
  intelligibility &  good (phone quality)  & bad (VHF quality + noise) \\ 
     \hline
    accents &  US English    & diverse \& non-native \\
    \hline 
   lexicon \& syntax & oral syntax, everyday topics  & limited to ICAO phraseology and related \\
   \hline
     speech rate &  standard  & high \\ 
     \hline 
     other  &  -  & code switching, possible Lombard effect  \\ 
    \end{tabular}
\end{table*}

In short, industrial stakeholders consider today that ASR and NLU technologies could help decrease operators' workload, both on pilots and on controllers' sides. A first step towards cognitive assistance in ATC-related tasks could be a system able to (1) provide a reliable transcription of an ATC message; and (2) identify automatically the call sign of the recipient aircraft.

Although significant progress has been made recently in the field of ASR --- see, for example, the work of \cite{Xiong_2016} and \cite{Saon_2017} who have both claimed to have reached human parity in the switchboard corpus \cite{Godfrey_1992} --- ATC communications still offer challenges to the ASR community; in particular because it combines several issues in speech recognition: accented speech, code-switching, bad audio quality, noisy environment, high speech rate and domain-specific language associated with a lack of voluminous datasets~\cite{Delpech_2018}. 
The Airbus Air Traffic Control Speech Recognition 2018 challenge was intended to provide the research community with an opportunity to address the specific issues of ATC speech recognition. 

This paper is an attempt to provide an overview on the challenge outcomes. Section~\ref{sec:format} presents the specificity of ATC speech as well as existing ATC speech corpora; section~\ref{sec:description} describes the tasks, dataset and evaluation metrics used in the challenge; section~\ref{sec:analysis} briefly describes the best performing systems and analyses the results of the challenge. Perspectives are discussed in section~\ref{sec:discussion}.

\begin{table}[hb]
\centering
\caption{Number of speech utterances and average duration within parentheses according to the speech program (AT: ATIS, AP: Approach, TO: Tower).} 
\label{tab:statcorpus}
\vspace{2mm}
\begin{tabular}{lccc}
    & ATIS &  AP &  TO \\
 \midrule
  train & 843 (27.6 s) & 20227 (4.5 s) & 6975 (4.3 s) \\
  dev & 102 (31.1 s) & \hphantom{1}2484 (4.3 s) & \hphantom{1}920 (4.2 s) \\
  test & 102 (30.4 s) & \hphantom{1}2600 (4.5 s) & \hphantom{1}893 (4.4 s) \\
\end{tabular}
\end{table}

\section{Specificity of ATC speech and existing ATC speech corpora}
\label{sec:format}

ATC communications being very specific, voluminous generic datasets like the SWITCHBOARD corpus \cite{Godfrey_1992}  cannot be used to build an ATC speech recognition system. Table \ref{tab:atc_vs_swb} provides a comparison of ATC speech vs. SWITCHBOARD speech. ATC speech provides many challenges to automatic speech recognition: audio quality is bad (VHF), the language is English but pronounced by non-native speakers, speech rate is higher than in CTS \cite{Cauldwell_2007} and there is also a lot of code switching. The only advantage of ATC compared to CTS is that the vocabulary is limited to the International Civil Aviation Organisation (ICAO) phraseology~\cite{ICAO_2007}. 

Several ATC datasets have been collected in the past. Unfortunately most of them are either unavailable, lack challenging features of ATC or lack proper annotation. On top of this, it was required that at least a small portion of the dataset had never been disclosed so that it could be used for evaluation.

The HIWIRE database~\cite{Segura_2007} contains military ATC-related voice commands uttered by non-native speakers and recorded in artificial conditions. The nnMTAC corpus~\cite{Pigeon_2007} contains 24h of real-life, non-native military ATC messages. Unfortunately, it is not available outside of NATO\footnote{North Atlantic Treaty Organization} groups and affiliates. Similarly, the VOCALISE dataset \cite{Graglia_2005} and the corpus of \cite{Lopez_2013} (respectively 150h and 22h of real-life French-accented civil ATC communications) are not publicly available. ATCOSIM ~\cite{Hofbauer_2008} is a freely available resource composed of realistic simulated ATC communications. 
Its limitations are its size (11h) and the fact that it lacks real-life features. The NIST Air Traffic Control Corpus \cite{Godfrey_1994} is composed of 70h of real-life ATC from 3 different US airports and it is commercially available through the Linguistic Data Consortium (LDC). Unfortunately, it is mainly composed of native English and the call signs have not been annotated. The corpus collected by \cite{Smidl_2014} is freely available and contains real-life non-native ATC speech. It is though quite small (20h) and does not contain call sign annotations.

\begin{figure*}[htb]
\begin{minipage}[b]{.48\linewidth}
 \centering
 \centerline{\includegraphics[scale=0.23]{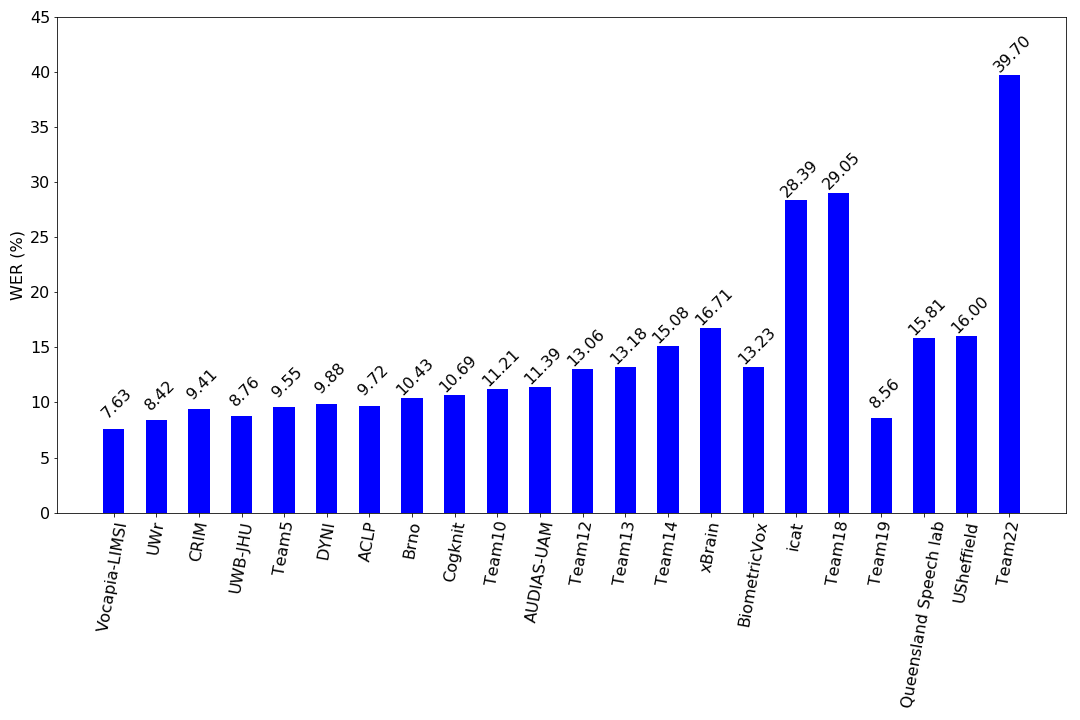}}
 \centerline{(a) ASR performance in Word Error Rates on Eval (\%)}\medskip
\end{minipage}
\hfill
\begin{minipage}[b]{0.48\linewidth}
 \centering
 \centerline{\includegraphics[scale=0.23]{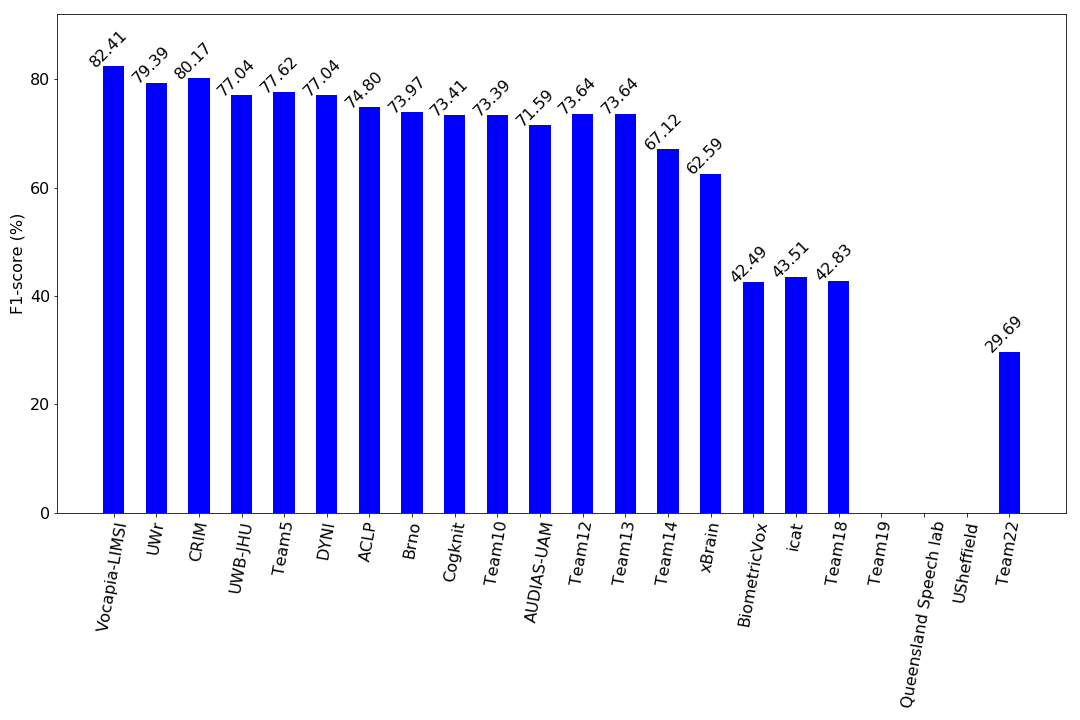}}
 \centerline{(b) CSD Performance in F1-score on Eval (\%).}\medskip
\end{minipage}
\caption{Performance of the 22 systems on the Eval subset. }
\label{fig:res}
\end{figure*}

\begin{table*}[htb]
\centering
\caption{Results for the ASR and CSD tasks for the five best ranked teams.} \label{tab:res}
\vspace{2mm}
\begin{tabular}{l|*{4}{c}|*{3}{c}}
            & \multicolumn{4}{c|}{ASR} & \multicolumn{3}{c}{CSD}\\ 
Team        & WER (\%)  & ins (\%)  & del (\%)  & sub (\%)  & F1 (\%)       & p (\%)     & r (\%)    \\ \midrule
 Vocapia-LIMSI   & \hphantom{1}7.62       & 1.29 & 3.14 & 3.19 & 82.41    & 81.99       & 82.82    \\
 UWr  & \hphantom{1}8.42       & 1.52 & 3.03 & 3.87 & 79.39    & 81.00      &   77.84    \\
 CRIM   & \hphantom{1}9.41       & 1.21 & 4.51 & 3.69 & 80.17    & 84.94      &    75.91   \\
 UWB-JHU  & \hphantom{1}8.76       & 1.55 & 3.42 & 3.80 & 77.04    & 84.05      &  71.11      \\
  Team5 & \hphantom{1}9.55       & 1.80 & 3.97 & 3.79 & 77.62    &  84.27     &  71.94     \\
\end{tabular}
\end{table*}

\section{Challenge description}
\label{sec:description}

\subsection{Two tasks: ASR and call sign detection (CSD)}

The Airbus ATC challenge consisted in tackling two tasks: 1) automatic speech-to-text  transcription from authentic recordings in accented English, 2) call sign detection (CSD). 

Aviation call signs (CS) are communication call signs assigned as unique identifiers to aircraft. They are expected to adhere to the following pre-defined format: an airline code followed by three to five numbers and zero to two letters. For instance, "ENAC School six seven november" is a call sign in which ENAC school is a company name followed by two numbers (six and seven) and "november" stands for the 'n' character in the aviation alphabet. One difficulty lies in the use of shortened spoken CS when there is no ambiguity.

\subsection{Speech material}


The dataset used for running the challenge is a subset of the transcribed ATC speech corpus collected by Airbus~\cite{Delpech_2018}. This corpus contains speech signals at 16 kHz sampling rate and 16 bits per sample. All the specific features of ATC mentioned above are included in the corpus: non-native speech, bad audio quality, code-switching, high speech rate, etc. On top of this, call signs contained in the audio have been tagged, which allowed the challenge organizers to propose a "call sign detection" task. Although the corpus is not publicly available, a subset of it was made available to the challengers, for challenge use only. Half of the whole corpus, totalling 50 hours of manually transcribed speech, was used. Utterances were isolated, randomly selected and shuffled. All the meta-information (speaker accent, role, timestamps, category of control) were removed. The corpus was then split into three different subsets: 40h of speech together with transcriptions and call sign tags for training, 5h of speech recordings for development (leaderboard) and 5h for final evaluation, were provided to the participants at different moments during the challenge. The participants did not have access to the ground-truth of the development and eval subsets. They could make submissions to a leaderboard to get their scores on the dev subset. Several criteria were considered to split the data into subsets that share similar characteristics (percentages given in speech duration): 1) speaker sex (female: 25\%, male: 75\%), 2) speaker job --- ATIS (Airline Travel Information System, mostly weather forecasts, 3\%), pilots  (54\%) and controllers (43\%) ---, the "program" --- ATIS (3\%), approach (72\%), tower (25\%). Table~\ref{tab:statcorpus} shows the number of utterances according to the program and the average mean duration of the utterances. ATIS is characterized by utterances of about 30~s in average longer than AP and TO with 4.5 second utterances in average.



Links to the other available ATC datasets (\cite{Hofbauer_2008,Godfrey_1994,Smidl_2014}) were given to the challengers so that they could use them as additional training data. Some participants did try to use external data with no gains or even with performance drops.

\subsection{Evaluation metrics}
Evaluation was performed on both the ASR and CSD tasks. ASR was evaluated with Word Error Rate (WER). Before comparison, hypothesis and reference texts were set to lower case. These are compared through dynamic programming with equal weights for deletions, insertions and substitutions. For CSD, F-measure ($F1$ or F1-score) was used. A score $S_{i}$ of a submission $i$ was defined to combine WER and F1 as the harmonic mean of the normalized pseudo-accuracy ($pACC_{i_{norm}}$) and the normalized $F1$ score ($F1_{i_{norm}}$):
$$ S_i = \frac{2\times pACC_{i_{norm}} \times F1_{i_{norm}}}{pACC_{i_{norm}}+F1_{i_{norm}}}$$ 

where
$$ pACC_i = 1 - min(1, \textnormal{WER}_i) $$
$$ \vec{v} : \text{submissions' scores vector}$$
$$ \vec{v}_{i_{norm}} = \frac{\vec{v}_{i} - min(\vec{v})}{max(\vec{v}) - min(\vec{v})}$$

The harmonic mean was chosen since it penalizes more strongly than the arithmetic mean situations where one of the two scores is low. Submissions were sorted by decreasing $S$ score values to get the final participant ranking.


\section{Result analysis and system overview}
\label{sec:analysis}

In this section, we report detailed results for the two tasks ASR and CSD. We also give a bird's eye view on the approaches of the best ranked predictions on the Eval subset. 

\subsection{Results}

Figures \ref{fig:res}a and \ref{fig:res}b show the Word Error Rates (WER) for the ASR task and the F1-scores for CSD, obtained by the 22 teams ordered by their final ranking. Only the names of the entities that gave a disclosure agreement are displayed. 

VOCAPIA-LIMSI achieved the best results in both tasks with a 7.62\% WER and a 82.41\% CSD F1-score. Globally speaking, the best teams obtained impressive results with WERs below 10\% and below 8\% for the winner. Table \ref{tab:res} gives more details to analyze these results. One can see that almost all the ASR systems produced twice as many deletions and substitutions (around 3\%) than insertions (around 1.5\%). 

Regarding CSD, the best systems yielded F1-score above 80\%. Except for the two best systems with similar precision and recall values (respectively 81.99\% and 82.82\% for VOCAPIA-LIMSI), precision was larger than recall by a significant margin. This means that the number of missed CS is larger than false alarms for these systems. This lack of robustness may be explained by the variability with which call signs are employed: sometimes in their full form, sometimes in partial forms. Three teams including Queensland Speech Lab and U. Sheffield did not submit CS predictions resulting in a zero score in CSD (no visible bar in fig. \ref{fig:res}b), and a final ranking that does not reflect their good performance in ASR.

\begin{table*}[ht]
\scriptsize
\centering
\renewcommand\thetable{5} 
\caption{Characteristics of the five best ranked teams' ASR systems.} \label{tab:models}
\vspace{2mm}
\begin{tabular}{l|*{9}{c}}
            & \multicolumn{2}{c|}{Acoustic frontend}& \multicolumn{3}{c|}{Acoustic Modeling}           & \multicolumn{2}{c|}{Language Modeling}       &   &       \\ 
    Team & Features  &  \multicolumn{1}{c|}{Data augmentation}  & Modeling  & Context   & \multicolumn{1}{c|}{Complexity}    & Lex. size  &\multicolumn{1}{c|}{ LM}        &     Decoding      &    Ensemble           \\ \midrule

Vocapia-LIMSI   & PLP-RASTA & No                        & HMM-MLP   & triphones & 6M            & 2.2k          & 4-gram            & Consensus & No    \\
UWr-ToopLoox    & Mel F-BANK& freq. shifting, noise    & CTC Conv-BiLSTM& diphones & 50M   & 2.2k          & 4-gram            & Lattice & Yes     \\
CRIM            & MFCC, ivectors& noise                 & BiLSTM-TDNN & triphones  &17M       & 190k          & RNNLM           & N-best & Yes     \\
UWB-JHU         & MFCC, ivectors& volume, speed         & TDNN-F & triphones   & 20M       & 2.2k            & 3-gram            & Lattice & No      \\
Team5 & MFCC, ivectors& reverb, speed, volume       & TDNN   & triphones & 6M       & 2.7k             & 4-gram            & Lattice & No      \\
\end{tabular}
\end{table*}

\begin{table}[H]
\centering
\renewcommand\thetable{4} 
\caption{Best ASR and CSD results according to the speech program (AT: ATIS, AP: Approach, TO: Tower), the speaker job (C: controllers, P: Pilots) and sex (F: female, M: male).} \label{tab:res2}
\vspace{2mm}
\begin{tabular}{l|*{3}{c}|*{2}{c}|*{2}{c|}}
 & \multicolumn{3}{c|}{Program} & \multicolumn{2}{c|}{Speaker} & \multicolumn{2}{c|}{Sex} \\
 & AT & AP & TO & C & P & F & M \\
 \midrule
 WER & 5.1 & \hphantom{8}8.1 & \hphantom{8}7.8 & \hphantom{8}5.5 & 10.5 & \hphantom{8}5.5 & \hphantom{8}8.2 \\
 F1 & \_ & 82.8 & 81.4 & 86.8 & 79.0 & 88.6 & 80.9 \\ 
\end{tabular}
\end{table}

To get more insights in these results, Table \ref{tab:res2} shows the highest ranked team WER and CSD F1-score according to the program, speaker job, and speaker sex. As expected, ATIS speech (mostly weather forecasts with limited vocabulary) is easier to transcribe than Approach (AP) and Tower (TO), for which similar WERs were obtained: 8.1\% and 7.8\%, respectively. An interesting finding is that pilots' speech (P) was much more difficult to transcribe than controllers' speech (C), with almost a factor two in WER, and 8\% absolute difference in CSD F1-score. This may be explained by the greater diversity of accents and speakers among pilots compared to controllers. Most of the controllers are French native speakers contrarily to the pilots. This could explain the better performance for controllers since French-accented English is the most represented accent in the corpus. Better performance was obtained for female speakers compared to male speakers probably because 78\% of the female utterances are controller utterances. This is also inline with results from the literature, where lower WERs on female speech ranging from 0.7 to 7\% were achieved depending on speech type condition~\cite{adda2005speech}.

\subsection{ASR system characteristics}

Table \ref{tab:models} gives an overview of the ASR modules used by the five best ranked teams. Regarding acoustic front-end, Vocapia-LIMSI used Perceptual Linear Predictive (PLP) features with RASTA-fil\-te\-ring~\cite{hermansky1990perceptual,hermansky1994rasta}. Except UWr-ToopLoox that used Mel F-BANK coefficients, all the other participants used high-resolution MFCC (40 to 80 coefficients) and 100-d i-vectors. According to their findings, i-vectors bring very small gains. For acoustic modeling, Vocapia-LIMSI used a hybrid HMM-MLP model (Hidden Markov Models - Multi-Layer Perceptron). UWr-ToopLoox used an ensemble of six large models (50M parameters each), each comprised of two convolution layers, five bidirectional Long Short-Term Memory layers (Bi-LSTM) trained with the CTC (Connectionist Temporal Classification,~\cite{graves2006connectionist}) objective function. CRIM also combined six different models, three Bi-LSTM and three Time-Delay Neural Networks (TDNN~\cite{Waibel89-PRU} using Kaldi~\cite{Povey11_TKS})~\cite{gupta2018crim}. UWB-JHU used factorized TDNNs (TDNN-F, \cite{Povey2018}), which are TDNNs whose layers are compressed via Singular Value Decomposition. 

Regarding developments specific to ATC speech, we noticed the use of specific pronunciations for certain words: words that correspond to letters (Alfa for A, Quebec for Q, using the NATO phonetic alphabet), and other cases such as niner for nine, and tree for three, for instance. Non-English word sequences, mostly French words, were denoted '@' in the manual annotations. Some systems used a special token for non-English words such as '$<$foreign$>$' and others simply mapped them to an unknown token ('$<$UNK$>$'). 

Finally, almost all the teams used the 2.2k word-type vocabulary extracted from the challenge corpus. The participants reported no gains when using neural language models rather than \textit{n-}gram models.



\subsection{Call Sign Detection system characteristics}

For CSD, two main approaches were implemented: on the one hand grammar-based and regular expression (RE) methods, \textit{i.e.} knowledge-based methods, on the other hand machine learning models. The first type of models requires adaptation to capture production variants that do not strictly respect CS rules (pilots and controllers often shorten CS for example). The second one, namely neural networks, Consensus Network Search (CNS), n-grams, perform better in this evaluation but are not able to detect unseen CS. Vocapia-LIMSI combined both approaches (RE allowing full and partial CSD together with CNS) and achieved the highest scores. 


\section{Summary and discussion}
\label{sec:discussion}

In this paper, we reported and analyzed the outcomes of the first edition of the Airbus and partners' ATC ASR challenge. 
The best ranked team achieved a 7.62\% Word Error Rate and a 82.41\% callsign detection F1-score on a 5-hour evaluation subset. ATIS speech, consisting of mostly  weather  forecasts  with  limited  vocabulary, was shown to be easier to transcribe than Approach and Tower speech interactions. Transcribing pilots' speech was found to be twice as harder as controllers' speech. 

Some participants attempted to use external ATC speech data for semi-supervised acoustic model training, and it was revealed to be unsuccessful. This technique usually brings performance gains, such as in \cite{vsmidl2018semi}. This may be due to the fact that the eval subset is very close to the trained one so that adding external data just adds noise. This outcome reveals a robustness issue that needs to be addressed. A large-scale speech data collection is very much needed to solve ATC ASR. Several criteria should be considered for this data collection: diversity in the airports where speech is collected, diversity in foreign accents, acoustic devices used for ATC, among others. 

Regarding organizing a future challenge, using speech from different airports for training and testing purposes should be considered. This also would require systems with more generalization capabilities for the CSD task since most of the call signs would be unseen during training.

Furthermore, to be successful, the major players in the field should join forces for data collection but also to share the large costs needed to manually transcribe the recordings. Finally, much attention should be paid to legal aspects on data protection and privacy (in Europe, the recent General Data Protection Regulation).


\section{Acknowledgements}
The organizing team would like to thank all the participants. This work was partially funded by AIRBUS\footnote{\url{https://www.airbus.com/}}, \'Ecole Nationale de l'Aviation Civile (ENAC)\footnote{\url{http://www.enac.fr/en}}, Institut de Recherche en Informatique de Toulouse\footnote{\url{https://www.irit.fr/}} and SAFETY DATA-CFH\footnote{\url{http://www.safety-data.com/en/}}.


\bibliographystyle{IEEEtran}
\bibliography{strings,main}

\end{document}